\documentclass[preprint,12pt]{elsarticle}

\usepackage{amssymb}
\usepackage{amsmath}
\usepackage{color}

\journal{Journal of High Energy Astrophysics}

\begin{document}

\begin{frontmatter}



\title{A new method for studying the blazar variability on the shortest time scales and its application to S5~1803+784}

\author[label1]{Butuzova M. S.}
 \affiliation[label1]{organization={Crimean Astrophysical Observatory},
             city={Nauchny},
             postcode={298409},
             country={Russia}}

\author[label1]{Guseva V. A.}

\author[label1]{Gorbachev M. A.}

\author[label1]{Krivenko A. S.}

\author[label1]{Nazarov S. V.}

\begin{abstract}
We propose a new method for investigating the evolution of the properties of the blazar brightness variability on timescales from a few hours to a few days.
Its essence lies in detecting sequentially located time intervals along the entire light curve, within which it is possible to determine the characteristic time of variability using the structure function.
We applied this method to uniform data series lasting several days provided by the TESS mission for blazar S5~1803+784.
Then, we analyzed the found time parameters of variability coupled with the data of B-, V-, R-, and I-photometric observations.
A correlation was found between the amplitude and the characteristic time of variability. 
The relation of these values with the spectral index of radiation has not been revealed.
We conclude that the variability on a short time scale is formed due to the different Doppler factors for having different volume parts of the optical emitting region. At the same time, the radiation spectrum deflects slightly from the power-law.
\end{abstract}


\end{frontmatter}



\section{Introduction}\label{intro}
Blazars are active galactic nuclei whose relativistic jet moves at a small angle to the line of sight. Almost all observed emissions from blazars are generated in their jets and exhibit brightness variability caused by changes in the relativistic amplification coefficient (doppler factor) and variations in the physical parameters of emitting regions. 
Optical radiation is generated in the region located inside the first parsec in a blazar's jet. The defining scientific achievements of recent years are associated with this region. Namely, the detection of extreme brightness temperature by the ground-space interferometer RadioAstron \cite{Kardashev17} and the connection of high-energy neutrinos detected by ICE Cube and Baikal-GVD with blazars \cite{Plavin20, Baikal24}. Explaining these facts requires revising the processes with high-energy particles in an ultrarelativistic jet.

Blazars have violent variability of emission flux across the whole electromagnetic spectrum on time scales from hours to tens of years.
Due to the Earth's daily rotation, optical observations of a separate blazar are carried out either continuously at night or once a several days/week for several years (for some objects interrupting due to the Earth's orbital motion).
Variability on time scales from day to a few days is studied poorly.
Almost continuous observations of a single object for several days are performed within the framework of international cooperation, the most famous of which is WEBT (Whole Earth Blazar Telescope).
For example, authors \cite{Bhatta13, Bhatta16} obtained and analyzed the 110-hour blazar S5~0716+714 light curve.
Kepler mission provided new opportunities, according to which \cite{Edelson13} obtained an 181-day light curve of the BL~Lac object 2R~1926+42 with a time resolution of 30 minutes.
The TESS\footnote{https://exoplanets.nasa.gov/tess/} (Transiting Exoplanets Survey Satellite) mission has significantly more sky coverage \cite{Ricker15}. TESS observations within at least one sector, i.e., $\sim$27~days with different time resolutions, cover almost the entire celestial sphere.
Data from one or more TESS sectors were used to analyze the variability of S5~0716+714 \cite{Rait21S5}, S4~0954+65 \cite{Rait21S4}, and BL~Lac \cite{Weaver20}.
For OJ~287, data are available for three consecutive TESS sectors \cite{Kishore24}.
The longest data series can be obtained for objects located near the ecliptic poles since TESS observed these areas continuously for several sectors (up to 13) twice and since January 2024 began the next such cycle.
Therefore, there is a unique opportunity to determine the shortest characteristic time of variability, which can range from fractions of a day to several days, and trace its evolution over a long period.
By supplementing the temporal characteristics of variability with spectral ones, we can conclude about the variability mechanisms on a short time scale.
Previously, no such study was performed because of the limitations of a uniformly sampled data series at one night or the use of an entire long-term data series with high temporal resolution.
In the second case, different characteristic times of variability canceled each other's manifestation in the results.

One of the blazars for which there is long-term TESS data is S5~1803+784, the study of whose variability on time scales from several hours to several days is devoted to this study.

Blazar S5~1803+784 ($z=0.68$, \cite{Lawrence96}) is very variable \cite{Kun18, Nesci21, Priya22, Carrasco23, Agarwal22}. For example, in the optical range, the maximum recorded rate of brightness change was 0.22 magnitude per day at the stage of flare growth, which peaked at MJD~56859 \cite{Nesci21}.
In the data from May 2020 to June 2021, Agarwal et al. \cite{Agarwal22} found a bright flare lasting from MJD~59063.5 to 59120.5. The maximum and minimum detected R-band magnitudes were 13.62 and 15.89.
That is, the brightness of S5~1803+784 is comparable to a magnitude for which the variability amplitude of the Poisson noise on the light curve is 20$\%$ or more relative to the average value of the recorded flux\footnote{https://heasarc.gsfc.nasa.gov/docs/tess/observing-technical.html, first plot}.
Therefore, for such a sufficiently faint object, the light curves obtained by standard TESS tools are very noisy, and there is a need to implement a new TESS data processing algorithm. We describe the various light curve types constructed from TESS data and supplement them with the Zwicky Transient Facility (ZTF) survey results in  Section~\ref{sec:data}. Section~\ref{sec:NewMethod} contains a new method to study the temporal characteristics of variability. The results, discussion, and conclusions are in Sections~\ref{sec:Result}, \ref{sec:Discuss}, and \ref{sec:Concl}, respectively.

\section{Data}
\label{sec:data}

We test a new method for investigating the evolution of the variability time scale on data from 41 sectors (from July 24 to September 20, 2021) --- the first published TESS data for S5~1803+784. Since the TESS mission destination is exoplanet detection, not to observe faint blazars, some features in the data need to be taken into account.

\subsection{TESS light curves}

\begin{figure}
\centering
\includegraphics[width= 0.95\columnwidth]{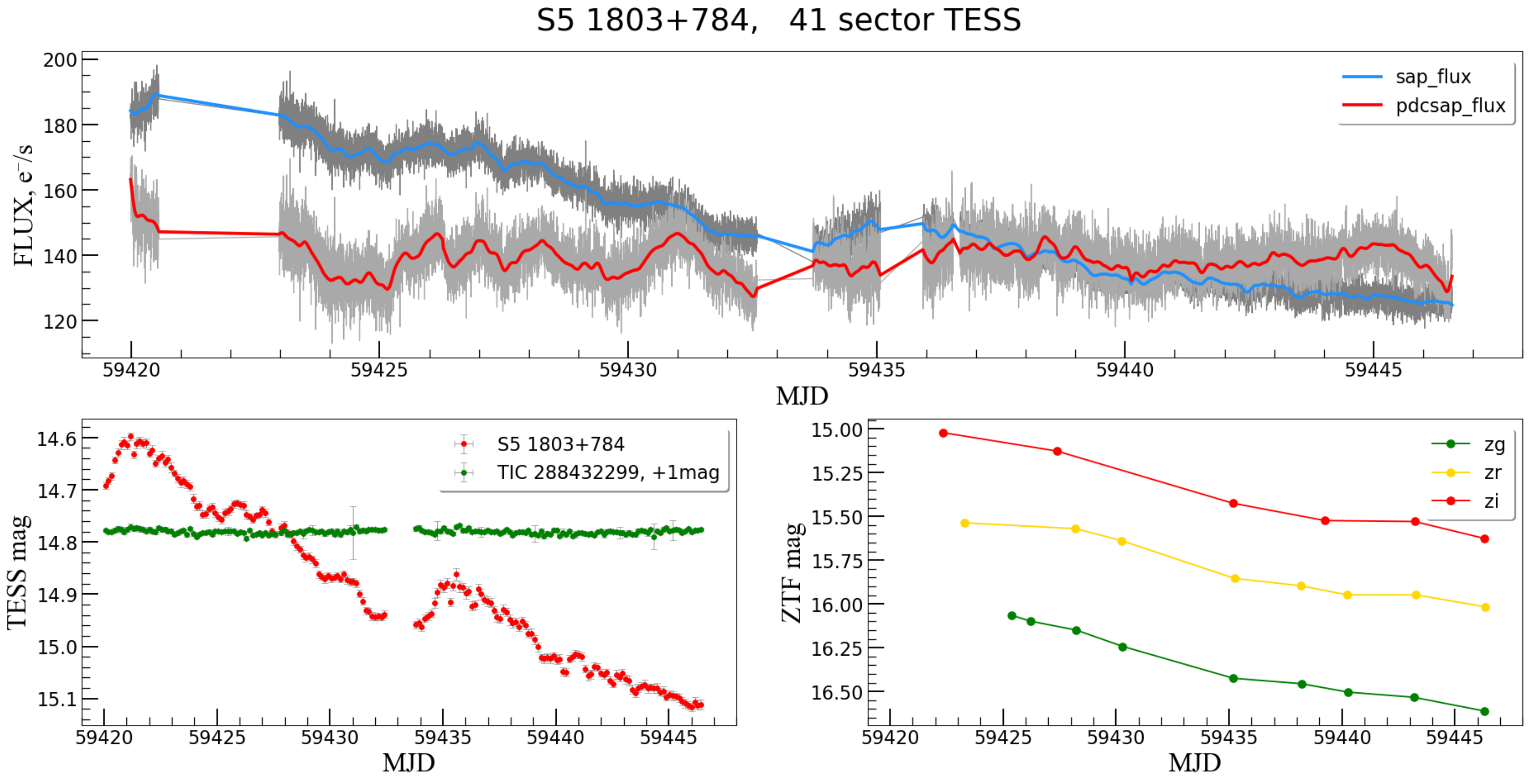}
\caption{The light curve S5~1803+784 from July 24 to August 20, 2021. Top panel: SAP and PDCSAP fluxes (grayscale) and their convolution with a Gaussian core of 75 points (blue and red lines, respectively). Bottom panel: on the left are the light curves obtained from aperture photometry of summarized cuts of TESS full-frame images for the object and the comparison star (red and green, respectively). On the right are the results of ZTF multiband photometry.}
\label{fig:fig_1}
\end{figure}

Using the Lightkurve\footnote{https://docs.lightkurve.org} package, we searched and extracted the object light curve of two types (Fig.~\ref{fig:fig_1}) by commands \textit{lk.search\_lightcurve()} and \textit{download()}. The first type is the so-called SAP flux (simple aperture photometry).
The SAP flux represents the total counts in target pixels (i.e., corresponding to the object), corrected for counts in background pixels.
The second type is a higher-level scientific product, the PDCSAP flux (Pre-search Data Conditioning SAP Flux). In PDCSAP flux, instrumental systematic trends are removed by excluding signals common to all objects in the field of view.
Since the primary goal of TESS is the detection of exoplanets, the trend accounting algorithms are configured in such a way as to simplify the detection of exoplanet transit in the light curves. This data processing method can remove the real astrophysical signal; therefore, we analyzed both data types.

Our study uses observational data with two- and ten-minute time resolution, which was processed by the Science Processing Operations Centre (SPOC) \cite{Jenkins16}.
It is worth noting that TESS CCD-detectors continuously read information at 2-second intervals, after which the received data is processed and summarized until the desired exposures are obtained. 
Thus, the TESS data represents a unique uniform long-term series.

There are no observations during periods when the TESS satellite transmits data to Earth or the spacecraft engines are running. Therefore, there are gaps in the uniform data series.
These data gaps can distort the result of detected changes in the characteristic time of variability ($\tau_\text{v}$) on the shortest time scales. 
To avoid this, we divided the SAP and PDCSAP light curves into parts, within which we allowed only the absence of a maximum of five points, located sequentially with a fixed time interval equal to 600 or 120 seconds.
As a result, we obtained four parts (Fig.~\ref{fig:fig_1}, upper panel), and for each of them, we separately applied the method described in Section~3. 
We excluded the part of MJD~59435.93 -- 59436.54 from consideration because $\tau_\text{v}$ was not to be found in any data within its boundaries. 
Note, on the considered time series, the gap of data points occurs twice: on the 2nd and 4th parts of the light curve and has a length of 5 points.

Due to the low brightness of the blazar S5~1803+784, the light curves in Fig.~\ref{fig:fig_1} are very noisy. To reduce the influence of the Poisson noise, we smoothed the light curves within each selected part through both the moving average method and Gaussian convolution with different window sizes. We used the window size in 5, 10, 15, 25, 50, 75, and, where possible, 125 points to further result comparison and develop criteria that determine the optimal window size for smoothing.

\subsection{TESS data aperture photometry}

\begin{figure}
\centering
\includegraphics[width=\columnwidth]{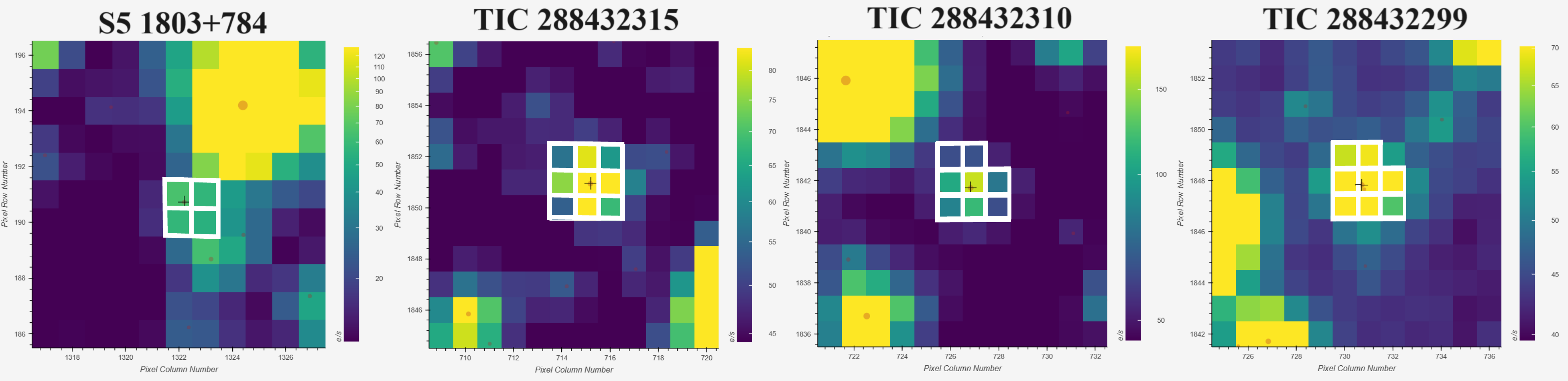}
\caption{Cuts from TESS full-frame images. The objects are indicated at the top of each plot, and their position is marked with a plus sign on the image. Red dots mark objects from Gaia~DR2. White borders denote the target pixels of the objects.}
\label{fig:fig_2}
\end{figure}

Another kind of TESS data is full-frame images, having a size of $12^\circ\times12^\circ$, with 1 pixel corresponding to $21^{\prime\prime}$. 
Faint sources occupy several pixels and merge with neighboring objects due to the low angular resolution in these images. Standard photometry programs, such as Iraf or MaxIm~DL, cannot work adequately with such images.
However, the object photometry relative to the comparison stars is necessary to restore the actual brightness change of the object. TESS procedures do not do this since their main aim is to identify the periodicity of stars caused by the presence of an exoplanet --- for which taking into account the trend on longer time scales is unnecessary. An additional difficulty in applying TESS data to the blazar study is the low brightness of the objects themselves.
The top panel of Fig.~\ref{fig:fig_1} shows that the light curves are strongly influenced by Poisson noise. The instrumental signal-to-noise ratio varies from 0.14 to 0.58, with a median of 0.31.
The MAST\footnote{https://archive.stsci.edu/} portal stores notes for each sector, including information on photometric accuracy, for the 41st\footnote{https://archive.stsci.edu/missions/tess/doc/tess\_drn/tess\_sector\_41\_drn59\_v02.pdf} sector too.
To implement aperture photometry of TESS data and improve the measurement accuracy of faint objects, we used cuts of $15\times13$ pixels from full-frame images, downloaded by query in https://mast.stsci.edu/tesscut/, for blazar S5~1803+784 and three comparison stars located at a distance of no more than $5^\prime$ (Fig.~\ref{fig:fig_2}).
The time resolution of these data series is 10 minutes.
Within the sections of a uniform data series for each source, we summarize a fixed number of cuts, in this case, 20.
All calculations were performed in the Mathematica Wolfram Research package. We used the command \textit{lk.interact\_sky()} in the Lightkurve package to display sources in the Gaia DR2 catalog to select the most optimal aperture for sources (Fig.~\ref{fig:fig_2}) and the number ($>30$) of pixels associated with the background.
Next, the following procedures were performed for each summarized image. We sorted the background pixel counts in ascending order and removed half of the pixels with large fluxes. Then, the script in the Mathematica Wolfram Research determined which random variable distribution the background corresponds to and created 100 model sets of background samples with the number of elements equal to the number of target pixels of the object.
After that, we subtracted the total background flux for each of the 100 sets from the total flux of the target pixels of the object. For the measured instrumental flux and its error, we took the median and the standard deviation from 100 measurements of the object flux. The accuracy of the determination for S5~1803+784 was from 0.5 to 1$\%$.
For comparison stars, the accuracy is higher than for the blazar since the stars themselves are brighter than the object. Using the magnitudes G in the Gaia band for comparison stars and transiting to the magnitudes T of the TESS filter according to the approximate formula $\text{T}=\text{G}-0.43$ \cite{Stassun18}, we obtained the light curve S5~1803+784 (Fig.~\ref{fig:fig_1}), which are in agreement with the SAP flux changes.

\subsection{ZTF data}

We used multiband photometric observations of ZTF in the considered time interval to trace the adequacy of the light curves obtained from TESS data.
For more information about the ZTF project, including one about the sensitivity of the devices, see \cite{Bellm19}. Fig.~\ref{fig:fig_1} shows that the blazar brightness gradually decreased continuously during the observation period.
The trend and amplitude of the changes agree with our photometry and the SAP light curve. It confirms the conclusion of Raiteri et al.~\cite{Rait21S5, Rait21S4} and Weaver et al.~\cite{Weaver20} that in PDCSAP fluxes, some removed long-term trends may have an astrophysical origin.

\begin{figure}
\centering
\includegraphics[width=0.6\columnwidth]{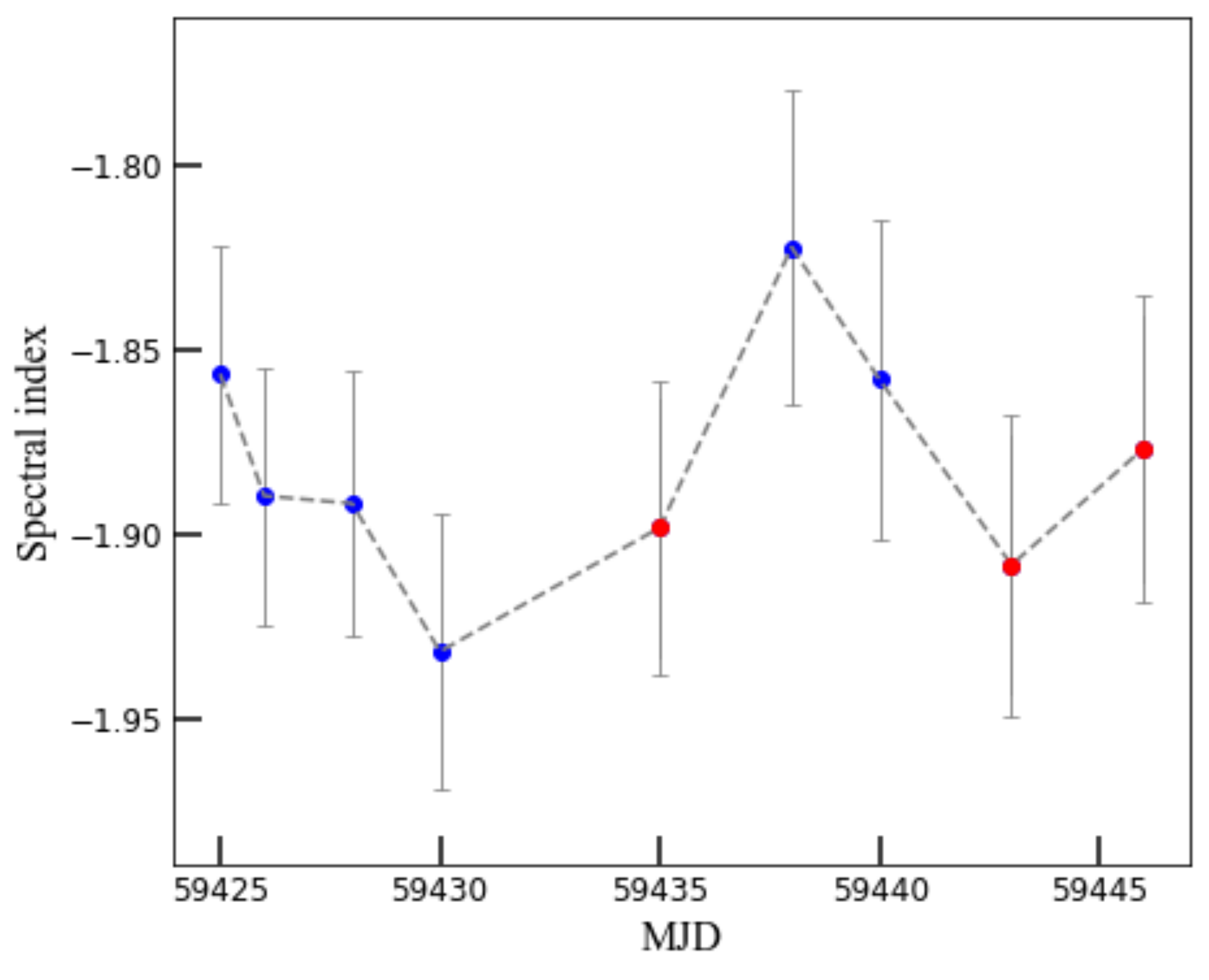}
\caption{Optical spectral index according to ZTF data from July 24 to September 20, 2021. The red and blue colors mark the points, for define of which data in three and two filters were used, respectively.}
\label{fig:fig_3}
\end{figure}

According to the data of quasi-simultaneous observations in 2 or more bands, we determined $\alpha$ within the framework of the power-law radiation spectrum $F\propto\nu^{\alpha}$ and linear approximation of the spectral flux density of radiation $F$ (from now on, we call them fluxes for short) for nearby dates (Fig.~\ref{fig:fig_3}).
The transition from magnitudes to fluxes was carried out by accounting for the ZTF filters corresponding to the AB system of magnitudes. The errors in spectral index calculations were estimated using the Python algorithm described by Gorbachev et al.~\cite{Gorbachev24}.

\section{A new method for determining the shortest characteristic time of variability}
\label{sec:NewMethod}

The brightness variability of blazars is often characterized as a superposition of oscillations with different time scales (see, e.g., \cite{Bhatta13,B21,Amir06}).
For research on the shortest time scales, we believe that at each moment, there is a single predominant process leading to the variability of the blazar's brightness, which either changes its parameters or is replaced by another one. In both cases, a change in the characteristic variability time is expected.

One of the traditional methods for determining the characteristic time of variability is to find the maximum of the structure function (SF) \cite{Sim85}, which is
\begin{equation}
   \text{SF}(\tau)=\frac{1}{N(\tau)}\sum^{N}_i \left[X(i+\tau)-X(\tau) \right]^2, 
   \label{eq:SF}
\end{equation}
where $X(i)$ and $X(i+\tau)$ are measurements at some point and after a time interval  $\tau$, respectively, $N(\tau)=\sum\left[\omega(i)\omega(i+\tau) \right]$ is the number of measurements, where the weight multiplier is $\omega=1$ if there is a measure and $\omega=0$ in the opposite case.
Therefore, the characteristic time of variability is $\tau$, at which SF has a maximum, and means the interval during which the change in the measured value will be maximum when registering changes in both directions.
Usually, the entire available light curve is used for SF analysis. However, even with a time resolution of the light curve of 600 seconds, there are 3169 flux measurements for 41 sectors selected according to the data quality parameter QUALITY=0.
(This condition means that there is no influence of various non-astrophysical effects in the data.) On such a long series, there may be a change in the characteristic time of variability $\tau_\text{v}$.
Therefore, the SF constructed for the entire series identifies only the dominant process, the time scale of which is not significantly less than the duration of the observation interval. At the same time, several consecutive processes with different $\tau_\text{v}$ are ``washed out'' in the resulting SF.
Additionally, long-term brightness changes are higher in amplitude than short-term variations. This fact also leads to the concealment of small characteristic variability times in the entire series analysis.

Short-term $\tau_\text{v}$ and their evolution could be detected by dividing the entire series into several intervals and constructing SFs for them. But in this case, there may be a situation where $\tau_\text{v}$ changes within the single interval.
To avoid this, we take the first 20 initial observation data points. At such an interval, it is unlikely that the SF will have a maximum. For this subsample, according to the formula (\ref{eq:SF}), we calculate SF in increments of $\tau$ equal to the time resolution of the series. These are 120 and 600 seconds for SAP and PDCSAP fluxes and 200 minutes for aperture photometry.
We check the presence of the maximum as follows. We select points on the SF whose ordinal number $i$ is within $i>3$ and $i<n-3$, where $n$ is the number of points in the SF.
For each selected point, we create two arrays. In the first one, we include the selected point and three SF points located on the left side, and for the second one --- the selected point and three points on the right side. Then we construct linear approximations for these arrays and determine the slopes.
We consider the maximum to be found if, firstly, for one or more consecutive points, the slope coefficient of the approximation on the left is positive, and on the right is negative.
Secondly, the peak SF value should be $\geqslant 0.0009$, corresponding to a brightness change of 0.03 magnitude during $\tau_\text{v}$ to cut off false SF maxima caused by receiver noise.
For the SF values to reflect the average square of the magnitude difference with the fixed time difference between measurements, we used $X(i)=\lg[X(i)]=\lg[F(t)]$ and inserted a multiplier of 2.5 to the difference of the measured values in the formula~(\ref{eq:SF}). The results of aperture photometry were substituted into the formula (\ref{eq:SF}) without changes.

\begin{figure}
\centering
\includegraphics[width=\columnwidth]{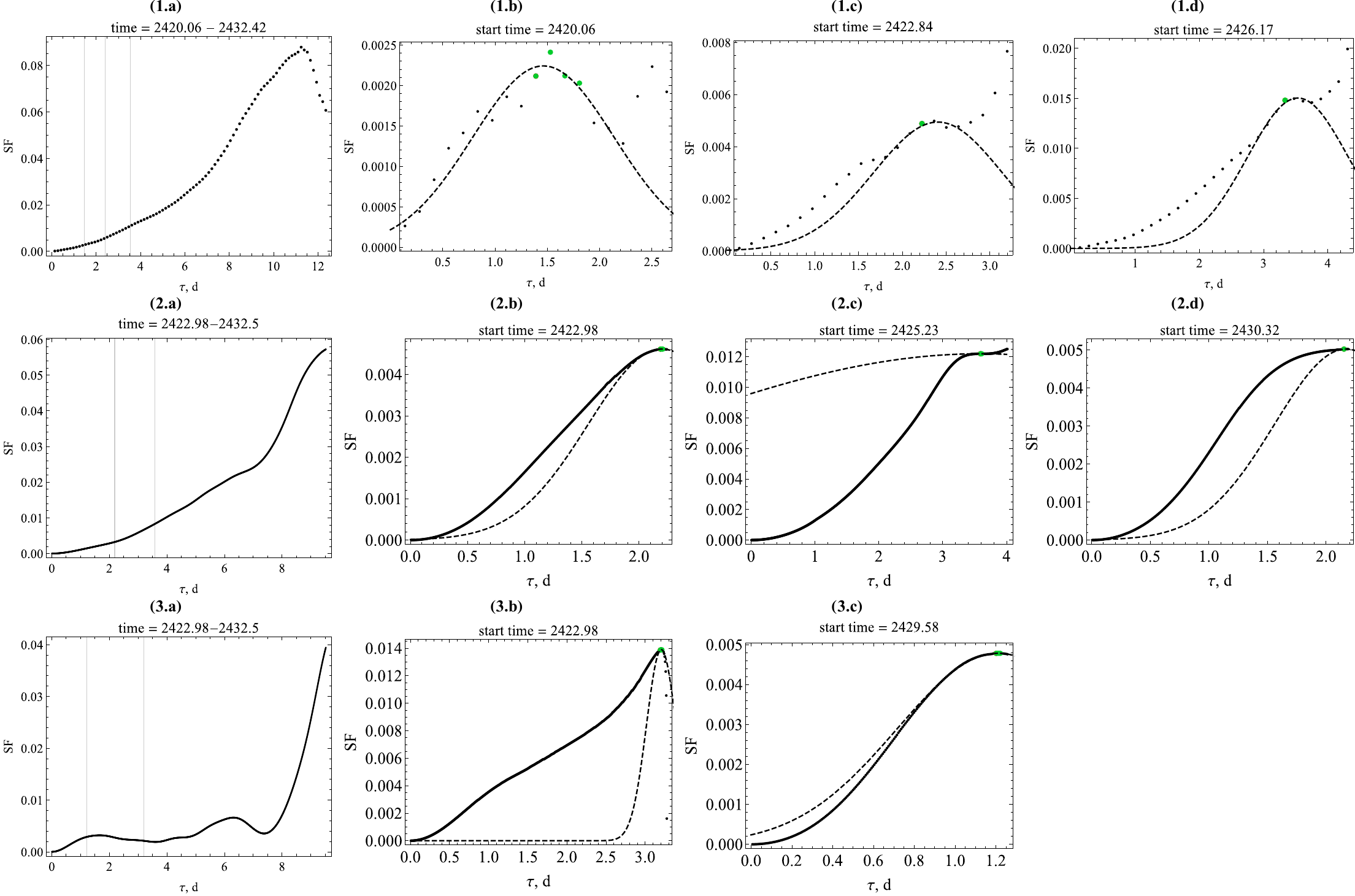}
\caption{Structure functions for a long continuous series of observations (on the left) and intervals isolated using a new method and characterized by a different $\tau_\text{v}$ (on the right).
The latter's plots show the Gaussians used to find the maximum (dotted line). Used for plotting data are (from top to bottom) aperture photometry, SAP and PDCSAP fluxes smoothed with a Gaussian core of 75 points.
The found $\tau_\text{v}$s for each data type are marked with vertical lines on the corresponding plots of the overall SF. For plots in the left column, the duration of the analyzed data series is indicated at the top, and for the rest, it is declared the start of the considered interval. The time is MJD+57000.}
\label{fig:SFs}
\end{figure}

If the peak is not found, the data point subsequent in the light curve is added to the initial array, and the SF calculation and analysis for the presence of a maximum are performed again. This algorithm is repeated until the maximum in SF is found.
If the SF peak is represented by a single point, we added adjoining three points from two sides to it and performed a Gaussian approximation for this array.
The maximum value of the Gaussian and its offset relative to zero is used as the maximum of the structure function $\text{SF}_\text{max}$ and $\tau_\text{v}$.
We used Gaussian fitting for an automatic possibility to determine the SF maximum more accurately to avoid the influence of a possible point spread (see Fig. 4, plot 1.b) and the discrepancy between the programmatically selected point corresponding to the SF peak and the SF peak. As the errors of the obtained values, we adopted the level 1$\sigma$, calculated for parameters of Gaussian height and peak position.
It turned out that the errors range from tenths of a percent to several percent of the parameter value.
Since the Gaussian and SF do not have a valid physical connection, it is not reasonable to use the width at half maximum as an error for $\tau_\text{v}$.
Moreover, the time resolution of the series affects the Gaussian width for approximately the same-shaped SFs; compare, for example, plots 1.d and 2.c in Fig.~4. Therefore, we did not use the Gaussian width in the study, and it did not affect the result of determining the characteristic time of variability.
If several points represent the SF peak, the point with the maximum SF value is selected for the Gaussian approximation procedure described above.
After determining the SF peak, the corresponding observational data is discarded, and the following 20 data points are taken, for which the algorithm for calculating the SF and finding the maximum is repeated. As a result, we get the division of the entire light curve into intervals within which $\text{SF}_\text{max}$ and $\tau_\text{v}$ are defined.
Fig.~\ref{fig:SFs} illustrates the effect of a new method for detecting the shortest variability time scales for the three data types under consideration. It can be seen that the found $\tau_\text{v}$ do not appear in the SF constructed over the entire light curve.
Remarkably, there is an agreement in $\tau_\text{v}\approx 2.5$ and $3.5$ days in the data of aperture photometry and SAP fluxes. 
Since the aperture photometry data has values earlier than the SAP and PDCSAP light curves, the first interval with the detected characteristic time of variability (plot~1.b, Fig.~\ref{fig:SFs}) is missing in the SAP and PDCSAP data. There is good agreement in the SF shapes and values of $\text{SF}_\text{max}$ and $\tau_\text{v}$ for approximately the same time intervals in the aperture photometry and SAP data (Fig.~\ref{fig:SFs}, plots 1.c and 2.b, 1.d and 2.c) even under different time resolutions (0.14 d and 600 s for the first and second rows, respectively).
The characteristic time of variability for the time interval shown in plot~2.d (Fig.~\ref{fig:SFs}) was not revealed in the aperture photometry data because its fourth interval has a shorter duration than $\tau_\text{v}$. Note, that some defined SF maxima (for example, plot~2.c, Fig.~\ref{fig:SFs}) look like inflection points. It is due to the fact that for automatic verification of the SF peak existence, we use 7 neighboring points at the $\tau$ increment corresponding to the time resolution of the series, which is three orders of magnitude less than the interval duration.
The significant difference in SF for PDCSAP fluxes probably arises from the long-term trend elimination.

Emmanoulopoulos et al. \cite{Emman10} noted that if the SF maximum is at the boundaries of the considered interval, this maximum may be false. The estimation of the significance level of the found maxima was carried out as follows.
For each selected time interval, we considered the light curve as a random process and determined the parameters of this process by the tools of the Mathematica Wolfram Research package. The light curves were described by ARIMA and sometimes by ARMA processes.
Using the found parameters, 1000 model light curves were constructed, for which SF was calculated. Then, structure functions for models and observational data were normalized to their maximal value.
The peaks of the model SFs were identified and counted up the number $(N)$ of those that: 1) are in the range $\tau_\text{v}\pm0.05 \tau_\text{max}$ (where $\tau_\text{max}$ is the maximum delay value for the corresponding observed SF, having maximum at $\tau_\text{v}$); 2) the values are equal to or exceed the values of the peak of the observed SF.
Then, the significance level is $p=1-N/1000$. Note that the model SF may have several maxima, with a dense arrangement of which within 10$\%$ of the interval from the corresponding observed $\tau_\text{v}$ and values greater than $\text{SF}_\text{max}$, the significance level may be negative. It was realized in one case, for which we took $p=0$.

\section{Results}
\label{sec:Result}

TESS provides three types of data (see Section~\ref{sec:data}). There is no obvious decision on which kind of data to use in studying the variability of blazars and how to analyze data with a high level of Poisson noise. To answer the question, we analyzed all the TESS data types, including those with different time resolutions. Additionally, we use various methods and windows for smoothing SAP and PDCSAP light curves to eliminate the influence of Poisson noise. Analyzing the results for a wide parameter range will allow us to establish optimal parameters for smoothing highly noisy data.

We emphasize that the proposed method for seeking the shortest (relative to the data series time resolution) time scales of variability has a principal assumption of the action at some point in time of one variability process, which then changes its parameters or is replaced by another one with a different $\tau_\text{v}$ regardless of TESS data types.

\subsection{The average variability amplitude}

In simple assumption, it is natural to expect a dependency of $\text{SF}_\text{max}(\tau_\text{v})$. By definition, the values of a structure function can be interpreted as the square of the average variability amplitude over a fixed time interval.
Then, $\text{SF}_\text{max}$ is the highest value of the square of the average variability amplitude between two measurements, the time difference of which lies in the certain interval centered in $\tau_\text{v}$.
If the variability is formed by changes in the physical conditions in the jet, then the higher the change in amplitude, the longer the process leading to variability acts. On the other hand, the variability on short time scales may be due to an increase in the coefficient of relativistic radiation amplification -- the Doppler factor $\delta=\left[\Gamma(1-\beta\cos\theta)\right]^{-1}$ (where $\Gamma$ is the jet Lorentz factor, $\beta$ is the jet speed in units of the speed of light, $\theta$ is the angle between the velocity vector and the line of sight) for some part of the emitting region \cite{B21}.
Then, the higher the amplitude, the greater the $\delta$ and the more time compression in the observer's reference frame. That is, with a decrease in $\tau_\text{v}$, an increase in $\text{SF}_\text{max}$ occurs under a condition of approximately equal volumes of the emitting subcomponents.

\begin{figure}
\centering
\includegraphics[width=\columnwidth]{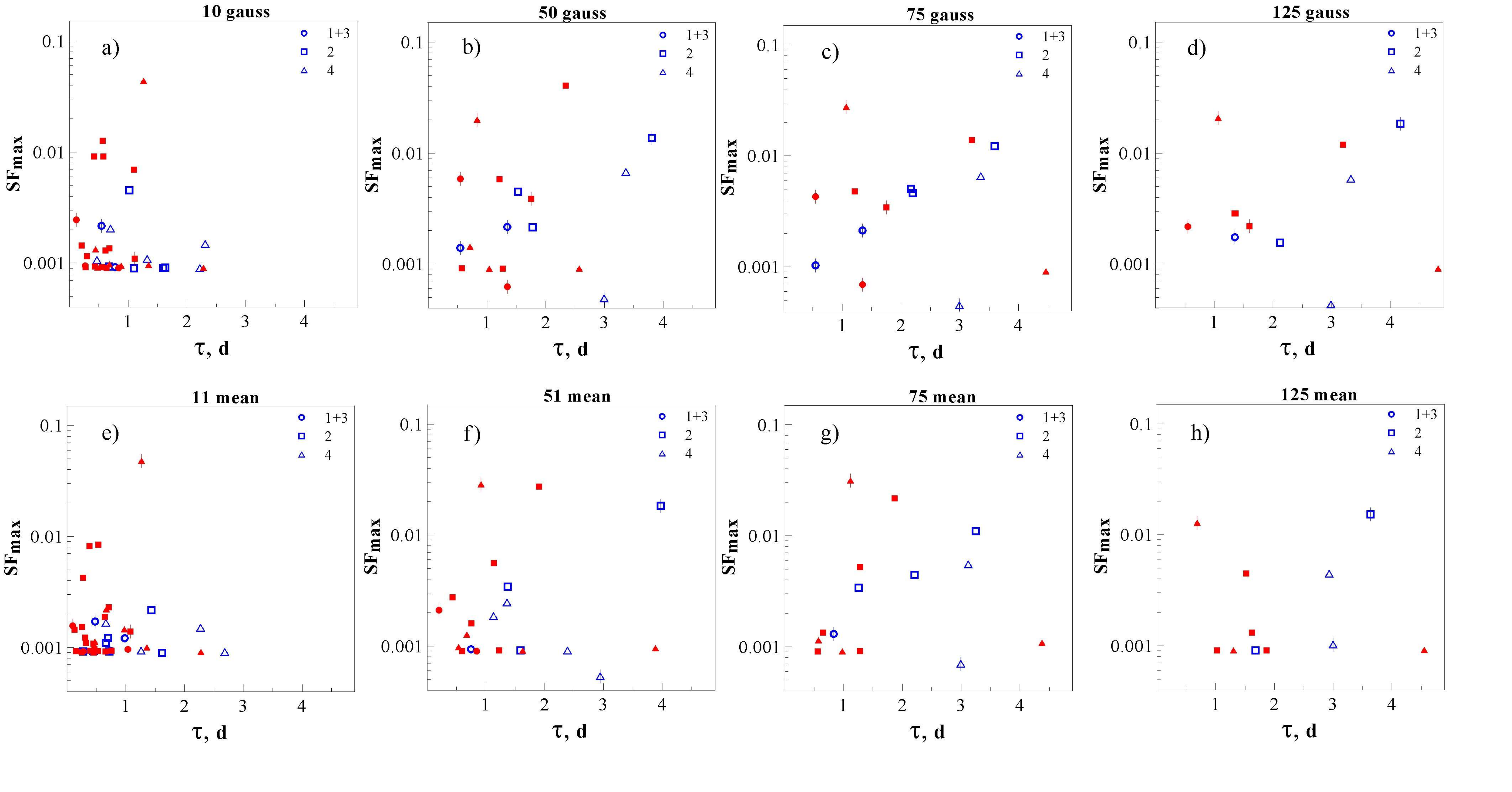}
\caption{ An illustration of the relation between $\text{SF}_\text{max}$ and $\tau_\text{v}$ for various types of blazar S5~1803+784 light curves according to TESS data for 41st sector with a resolution of 600 seconds. The upper panel shows the results of smoothing noisy data by convolution with Gaussian, the core width of which is indicated at the top of each plot; the lower panel represents results for smoothing by the moving average method. The points obtained based on SAP and PDCSAP fluxes are marked in blue and red, respectively. Sections of the light curve are marked with various symbols, explained in the plot legends. Symbols with a vertical line mark the maximum values of the delay $\tau$ and the increasing SF, the maximum of which is not determined due to the limitations of the data series. 
}
\label{fig:fig_5}
\end{figure}

\begin{figure}
\centering
\includegraphics[width=\columnwidth]{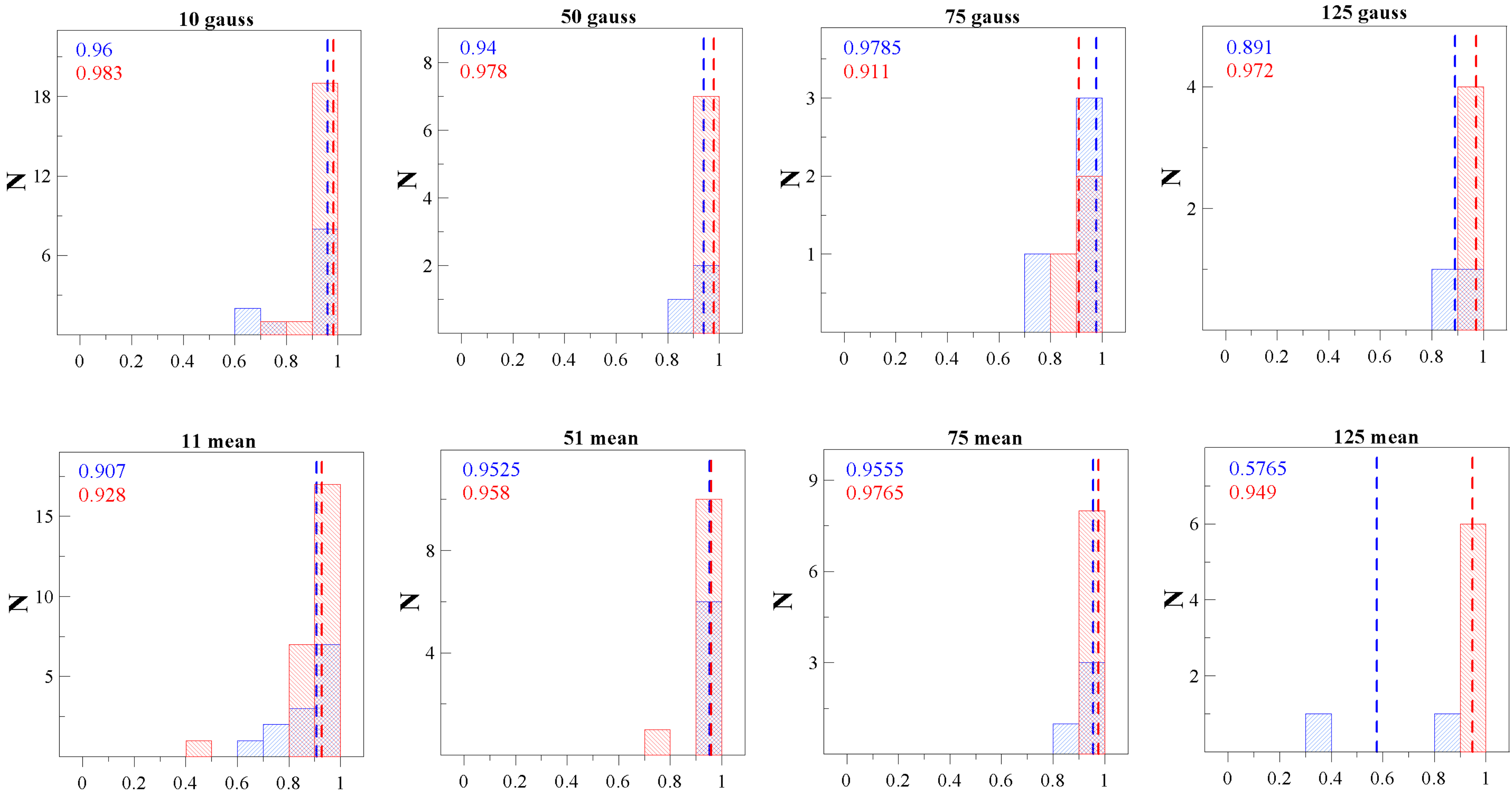}
\caption{Distribution of significance levels for the found SF peaks for data with a time resolution of 600 seconds. The results based on SAP and PDCSAP fluxes are shown in blue and red, respectively. The dotted line marks the median significance level.}
\label{fig:fig_6}
\end{figure}

The resulting dependencies $\text{SF}_\text{max}(\tau_\text{v})$ for data with a time resolution of 600 seconds are shown in Fig.~\ref{fig:fig_5}. 
Firstly, it is paying attention to the ``plateau'' at the level of the set minimum value of $\text{SF}_\text{max}$ at 0.0009 for cases of smoothing over less than 75 points.
We believe this situation is because many peaks with $\text{SF}_\text{max} \approx 0.0009$ represent random fluctuations of increasing SF, which satisfy the selected criteria for the SF maximum.
Under this, the more noisy the data, the greater the amplitude of these fluctuations and the higher the probability of false SF maxima. This conclusion is confirmed by the distributions of the significance level of the maxima (Fig.~\ref{fig:fig_6}).

Comparing the plots shown in Fig.~\ref{fig:fig_5}, it can be seen that in cases g) and h), there is still a ``plateau'' at the set minimum value of 0.0009, and in c) and d), there is not.
That is, the 75 points are already enough to smooth out the noisy light curve to eliminate the influence of noise on studies of the temporal variability characteristics. The approximate correspondence of some points for both SAP and PDCSAP fluxes in graphs c) and g)  evidences that the found $\tau_\text{v}$ reflects the real variability of the source.
Some points on one plot do not have an analog on the other, which probably occurs as a result of using various methods to smooth out the noisy signal.
On the plots in the two right columns of Fig.~\ref{fig:fig_5}, points for the SAP flux have a trend of increasing the value of $\text{SF}_\text{max}$ with an increase in the characteristic time of variability.
For points corresponding to the PDCSAP flux, the dependence $\text{SF}_\text{max}(\tau)$ is missing, and the points tend to be located on smaller $\tau$.
We believe that this is because due to the fact the long-term trend of brightness attenuation, presented in both ZTF data and aperture photometry, has been removed under calculating PDCSAP fluxes.
Additionally, plots in the first column of Fig.~5 contain points reflecting the high-amplitude fast variability in the PDCSAP light curve (their $\text{SF}_\text{max} \approx 0.01$ and $\tau_\text{v}\approx 0.5$~d). These points do not have counterparts in the SAP data and probably arise from the inability of small window smoothing to eliminate the effect of noise on the PDCSAP light curve. Thus, differences in the results obtained based on SAP and PDCSAP light curves arise due to algorithms for obtaining high-level scientific products such as PDCSAP-fluxes, which can increase the noise and remove both an instrumental signal and some astrophysical signal from SAP fluxes to simplify the analysis of the PDCSAP light curve for periodicity caused by an exoplanet \cite{Ricker15}.

When smoothing with a moving average over 75 and 125 points (plots g, h), the results differ significantly more than under convolution with Gaussian (plots c, d).
Moreover, some points are similar in c) and d), probably because Gaussian smoothing preserves small-scale fluctuations in highly noisy data better than a moving average.
From the comparison of the plots of the 3rd and 4th columns of Fig.~\ref{fig:fig_5}, there is a tendency for the points to shift towards an increase in the characteristic time of variability, which we associate with the ``blurring" of small-scale fluctuations with an increase in the points of the smoothing window.

\begin{figure}
\centering
\includegraphics[width=\columnwidth]{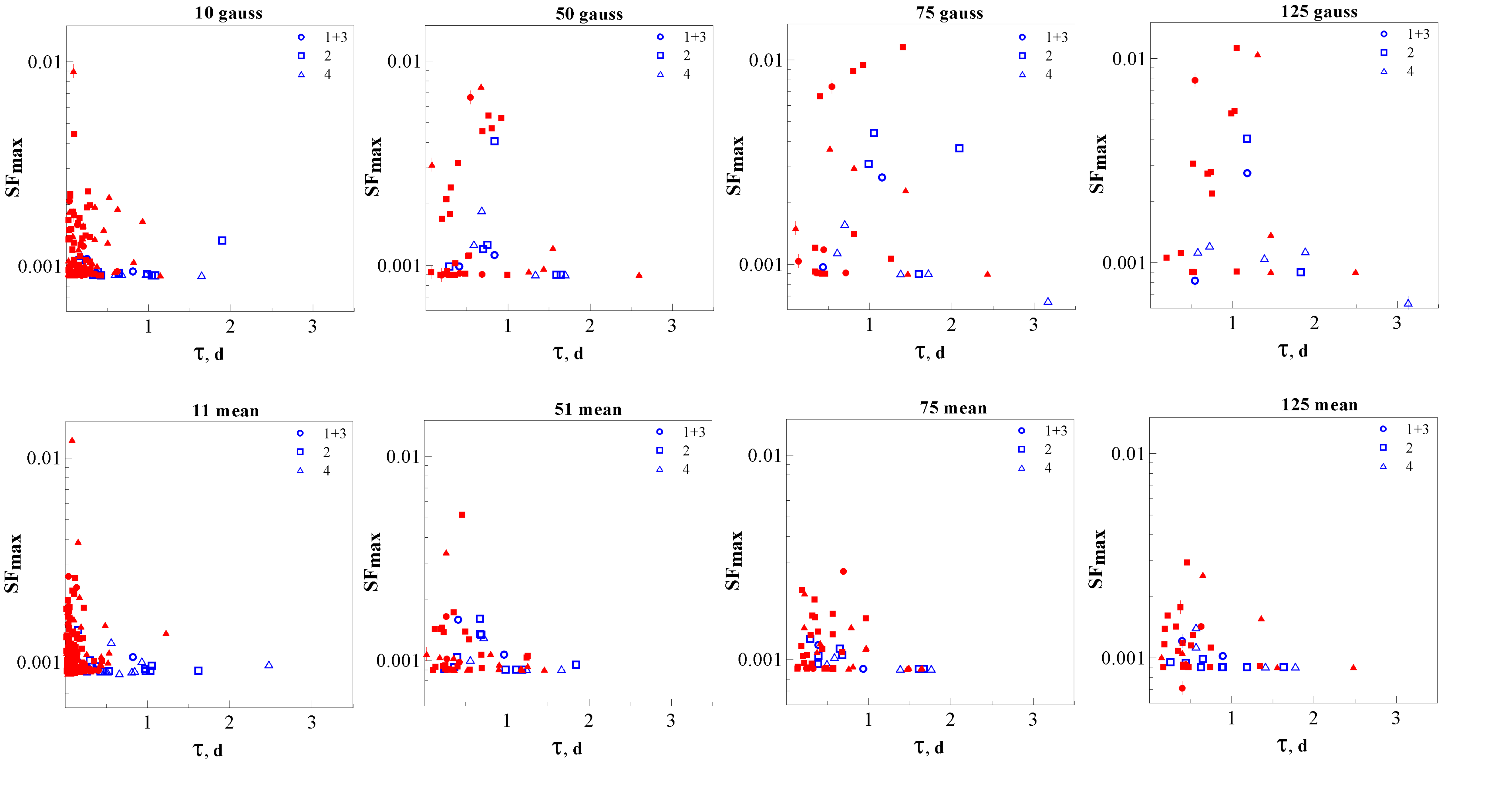}
\caption{An illustration of the relation between $\text{SF}_\text{max}$ and $\tau_\text{v}$ for various types of blazar S5~1803+784 light curves for 41st TESS sectors with a time resolution of 120 seconds. The designations are similar to Fig.~\ref{fig:fig_5}.}
\label{fig:fig_7}
\end{figure}

For data obtained with a time resolution of 120 seconds, the plateau at $\text{SF}_\text{max}=0.0009$ is maintained up to 125 point smoothing, the time interval for which is equal to the smoothing interval for 25 data points with a resolution of 600 seconds.
Therefore, further increasing the number of points used in smoothing 120-second data does not make sense.
Comparing the plots, corresponding to different methods of the original signal smoothing, it can be seen that convolution, with an increase in the number of points in the Gaussian core, leads faster to a shift of points from the lower limit of $\text{SF}_\text{max}$ towards increasing values than with smoothing by the moving average method (Fig.~\ref{fig:fig_7}).

\begin{figure}
\centering
\includegraphics[width=\columnwidth]{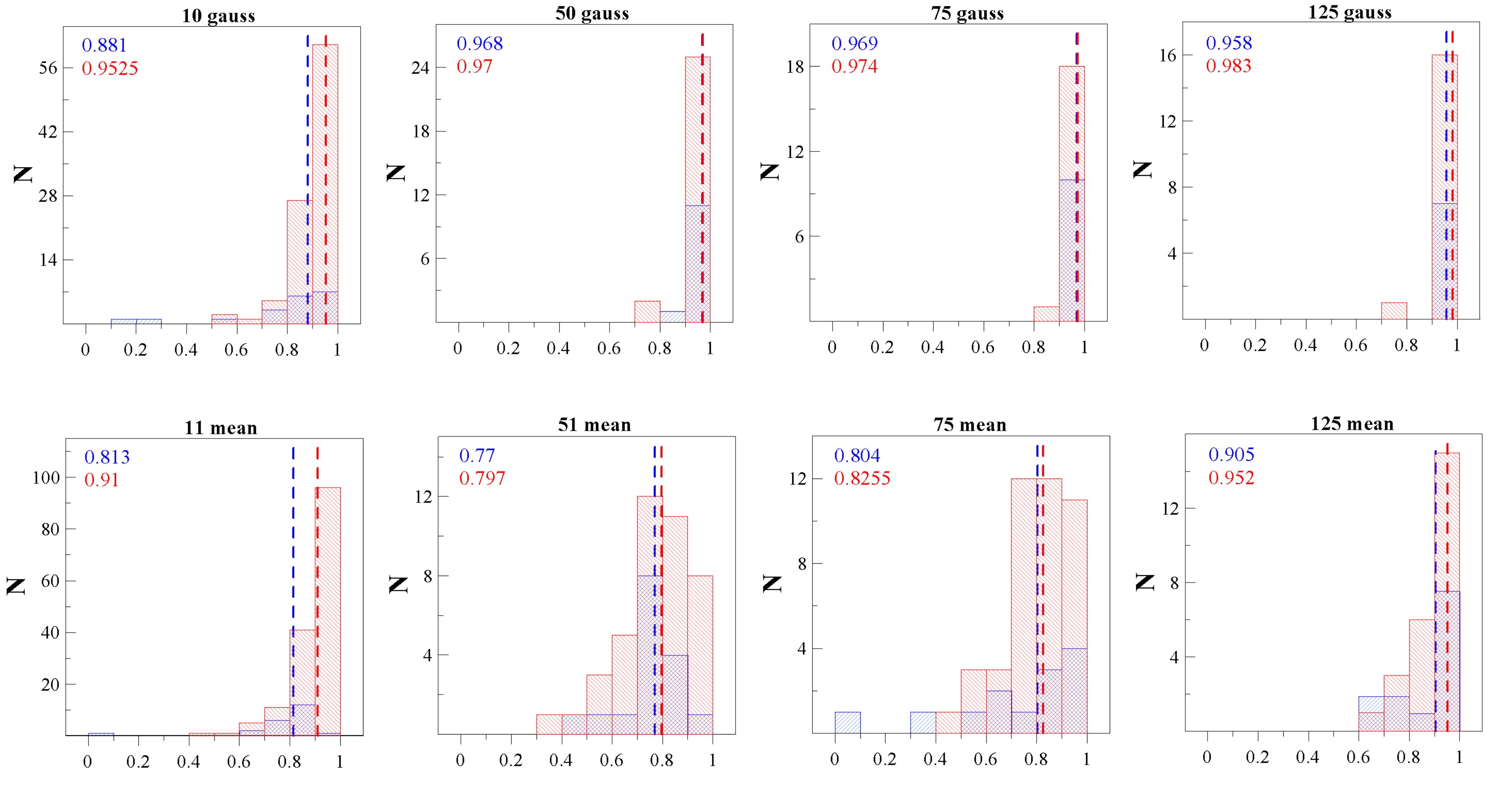}
\caption{Distribution of significance levels for the found SF peaks for data with a time resolution of 120 seconds. The designations are similar to Fig.~\ref{fig:fig_6}.}
\label{fig:fig_8}
\end{figure}

The distribution of the significance levels of the detected peaks in the structure functions (Fig.~\ref{fig:fig_8}) shifted to lower values relative to similar distributions for data with a time resolution of 600 seconds.  Therefore, the existence of a ``plateau'' in Fig.~\ref{fig:fig_7} confirms our conclusion about the strong influence of noise on the result of the variability analysis on the shortest time scales.

It is important to note that 120-second data were used to analyze the brightness micro-variability of blazars S5~0716+714 \cite{Rait21S5} and S4~0954+65 \cite{Rait21S4}. A break in the power spectrum of the light curve was detected for both sources.
This beak indicates that the variability on time scales less than $1-$1.5 hours has the character of white noise, whereas on higher time scales, it corresponds with red noise. However, the interval of $1-$1.5 hours is ten times less than the interval within which we performed smoothing to eliminate the strong influence of noise on the result.
Therefore, the detected break in the power spectrum \cite{Rait21S5,Rait21S4} is most likely due to the high noise level of the data. In the further analysis, we do not use a light curve with a time resolution of 120 seconds.

\subsection{Analysis of characteristic times of variability}

\begin{figure}
\centering
\includegraphics[width=\columnwidth]{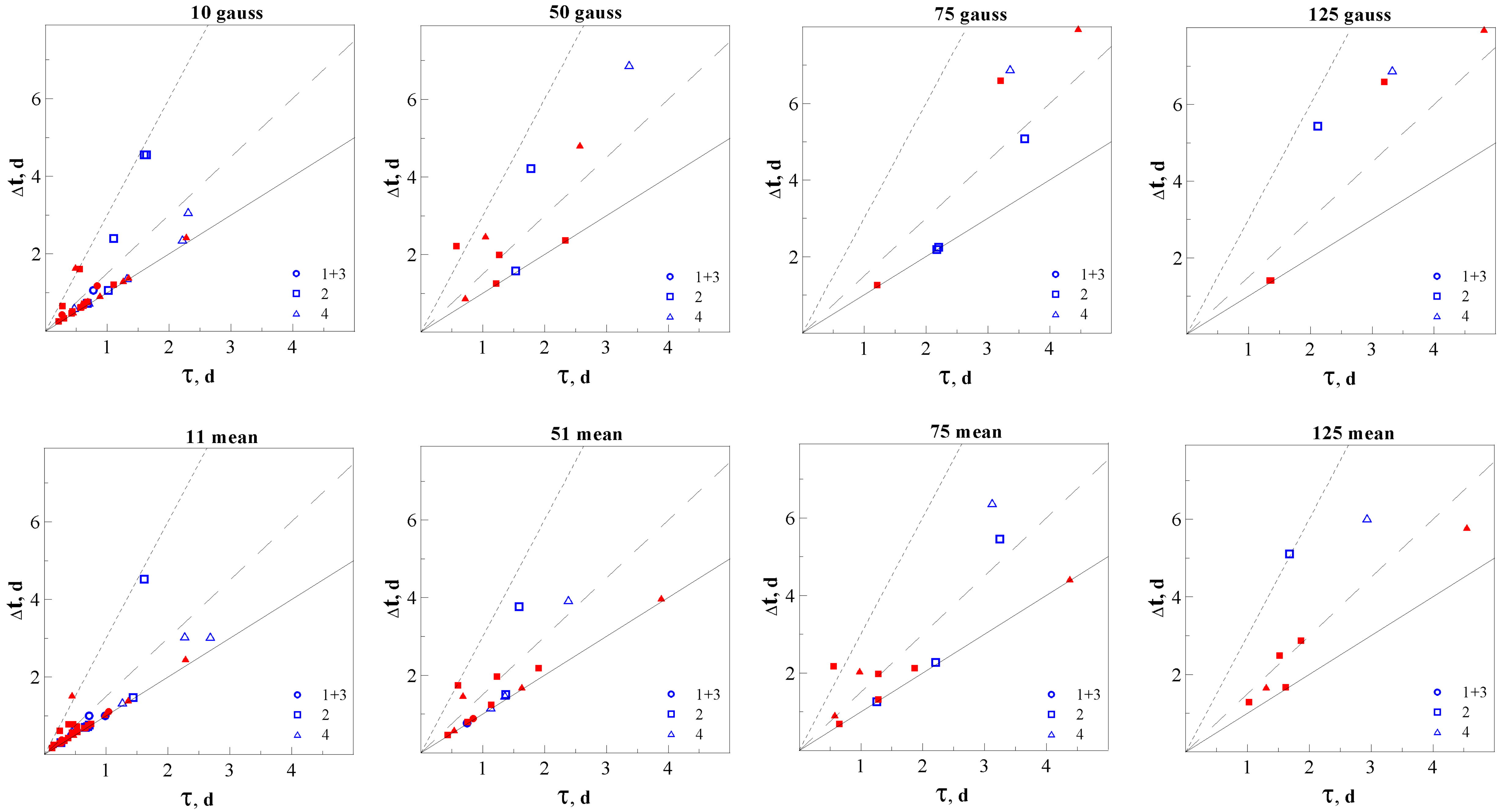}
\caption{The relation of the interval duration with the characteristic time of variability within it for light curves with the time resolution of 600 seconds. The designations are similar to Fig.~\ref{fig:fig_5}. The solid line marks $\tau=\Delta t$. The long and short dashed lines mark the characteristic time of variability in 2/3 and 1/3 of the length of the interval $\Delta t$.}
\label{fig:fig_9}
\end{figure}

\begin{figure}
\centering
\includegraphics[width=\columnwidth]{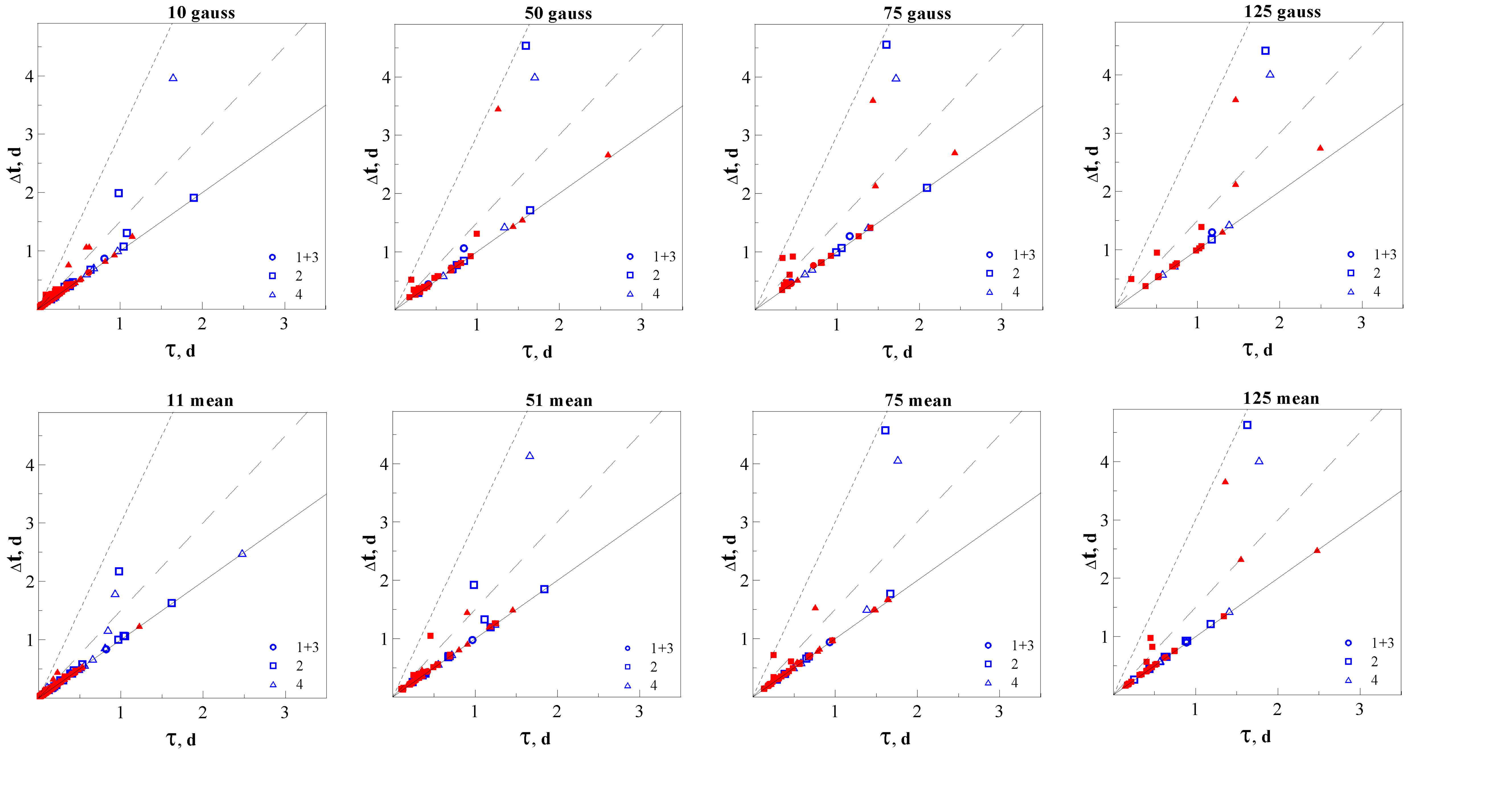}
\caption{The relation of the interval duration with the characteristic time of variability within it for light curves with a time resolution of 120 seconds. The designations are similar to Fig.~\ref{fig:fig_5}. The solid line marks $\tau=\Delta t$. The long and short dashed lines mark the characteristic time of variability in 2/3 and 1/3 of the length of the interval $\Delta t$.}
\label{fig:fig_10}
\end{figure}

The dependence of interval duration $\Delta t$ on $\tau$ (Fig.~\ref{fig:fig_9} and ~\ref{fig:fig_10}) also confirms the conclusion that with a small number of points used in smoothing a highly noisy signal, the detected characteristic times of variability are false due to random fluctuations in the structure function.
Namely, it can be seen that with a small number of points used for smoothing, the found characteristic times of variability are approximately equal to the duration of the intervals on which they are determined. As the points used for smoothing increase, the values of $\tau_\text{v}$ become noticeably shorter than $\Delta t$. 
For data with a time resolution of 120 seconds, even when smoothing with a window of 125 points, $\tau_\text{v}\approx\Delta t$ (Fig.~\ref{fig:fig_10}), which once again proves the impossibility of analyzing this data for a faint source with a flux of up to 200 counts/s.

Two reasons cause that the $\tau_\text{v}$ values range $\gtrsim \Delta t /3$. Firstly, the absence of high amplitude fluctuations on the time scale of the order of fractions of a day.
Secondly, the time series analysis algorithm is implemented in such a way that when $\tau_\text{v}$ is found (i.e., the maximum of SF is detected) in a certain time interval, a transition occurs to the next time interval, which begins with 20 points on the light curve and increases until the next $\tau_\text{v}$ will not be found.
Therefore, the found $\tau_\text{v}$ in a given time interval can characterize the variability for some fraction of the next time interval and be washed away by the dominant other characteristic time of variability.

\begin{figure}
\centering
\includegraphics[width=\columnwidth]{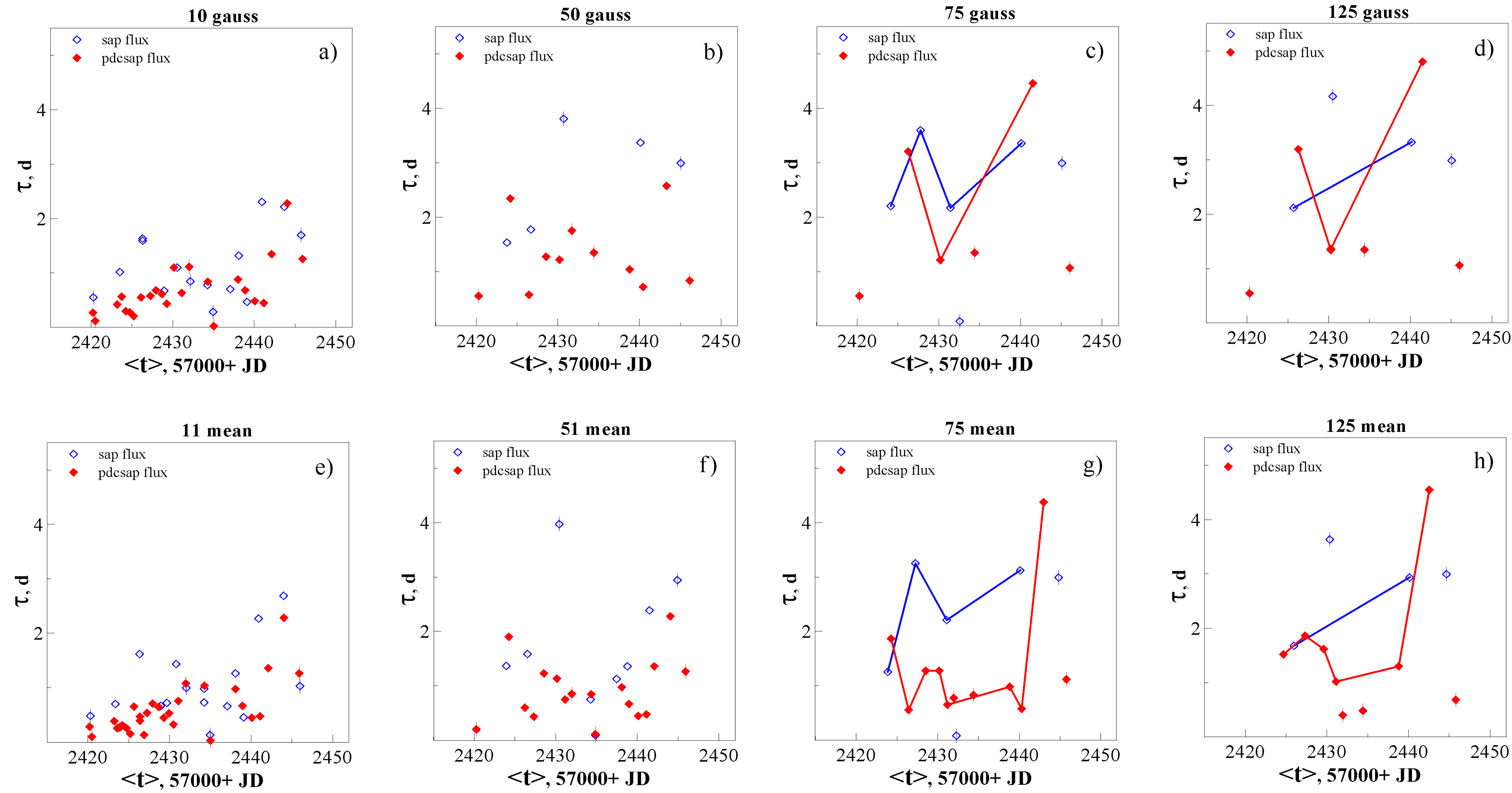}
\caption{Evolution of the characteristic time of variability. Except for the point symbols, the designations are similar to Fig.~\ref{fig:fig_5}. For the convenience of visual comparison of the plots of the third and fourth columns, the found $\tau_\text{v}$ are connected by lines.}
\label{fig:fig_11}
\end{figure}

Fig.~\ref{fig:fig_11} represents the change of $\tau_\text{v}$ over time. Let's first consider the points corresponding to SAP fluxes. The boundary points on plot c) agree well with similar points on plots d) and h).
The second and third points on plots c) and g) also agree. The rest distinction may be caused by a difference in the smoothing methods of the noisy light curve.
Plots c), d), g), and h) for data corresponding to PDCSAP fluxes show small $\tau_\text{v}$ in the middle and longer characteristic times of variability at the beginning and end of the 41st TESS sector. This trend is also evident in plots b) and f).

Thus, we register characteristic times of variability from $\approx$0.5 to 5 days. At adjacent time intervals, $\tau_\text{v}$ may differ significantly (almost 4 times).
The blazar S5~1803+784 light curve (Fig.~\ref{fig:fig_1}) shows either a gradual decrease in brightness (SAP fluxes) or an approximately constant level (PDCSAP fluxes). Therefore, the trends of $\tau_\text{v}$ (Fig.~\ref{fig:fig_11}) do not depend on the object's brightness.
Note that the 5-point gaps present on the light curve of the 2nd and 4th sections did not affect the result of calculations of $\text{SF}_\text{max}$ and $\tau_\text{v}$ since they are in the middle of the selected intervals in data with the 600-second resolution, convolved with a Gaussian of 75 points, and the duration of these intervals is several days.

\begin{figure}
\centering
\includegraphics[width=\columnwidth]{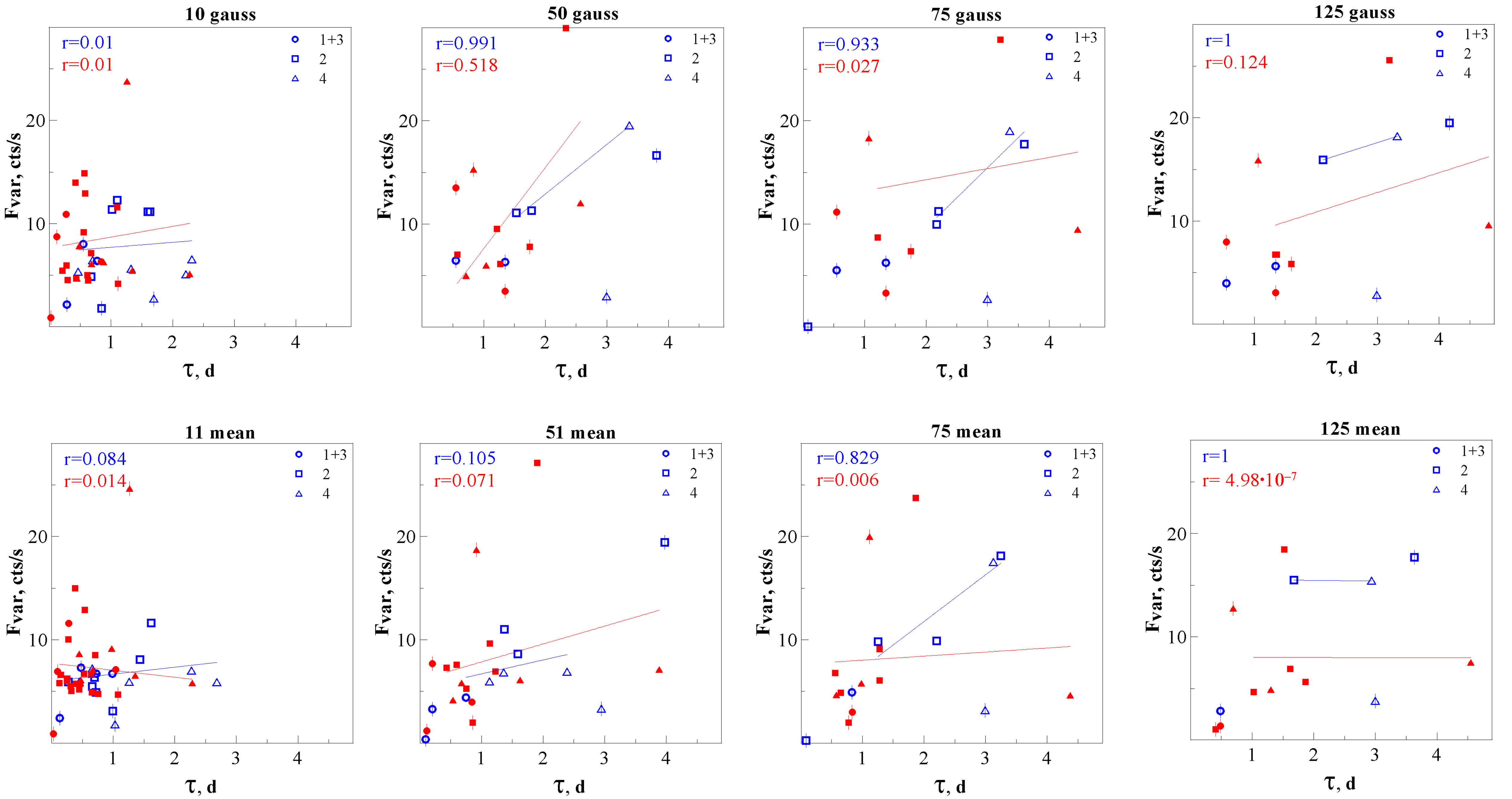}
\caption{The dependence of the maximum brightness change within the considered intervals of the light curve on the characteristic time of variability for data with a time resolution of 600 seconds. The designations are similar to Fig.~\ref{fig:fig_5}. At the top of each plot,  there is the corresponding Pearson correlation coefficient $r$, calculated using a logarithmic scale for the flux. The time is in units of MJD$-$57000.}
\label{fig:fig_12}
\end{figure}

With a smoothing window of 75 points for the SAP flux light curve, the dependence of the maximum variability amplitude within the considered intervals on $\tau_\text{v}$ appears (Fig.~\ref{fig:fig_12}).
On the other hand, there is no such correlation in the PDCSAP fluxes data. We believe this is because the maximum brightness change at the selected intervals is determined by a long-term trend, which is eliminated in the PDCSAP light curve.
When smoothing with a window of 125 points, there is no correlation either because of the small number of points or the ``blurring'' of low-amplitude fluctuations.

\subsection{$\text{SF}_\text{max}$ and $\tau_\text{v}$ based on aperture photometry}

\begin{figure}
\centering
\includegraphics[width=\columnwidth]{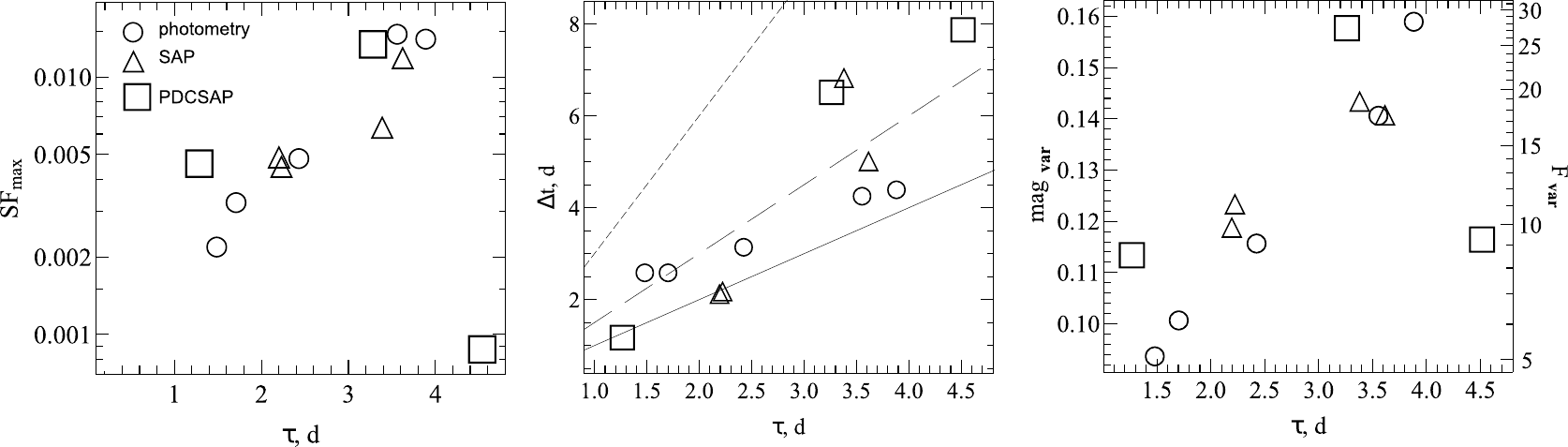}
\caption{The square of the average amplitude (on the left), the interval durations (in the center), and the maximum amplitude (on the right), depending on the characteristic time of variability obtained as a result of the analysis of various types of data. The solid line marks $\tau=\Delta t$. The long and short dashed lines denote the characteristic time of variability in 2/3 and 1/3 of the length of the interval $\Delta t$. PDCSAP and SAP data are given for smoothed Gaussians with a core of 75 points of the corresponding light curves.}
\label{fig:fig_13}
\end{figure}

We performed a similar analysis for the S5~1803+784 light curve, obtained by photometry of summed 20 TESS cuts of full-frame images (sec.~2.2).
We took the object's magnitude as $X(i)$ in Formula (1).
Note, that the minimum value of $\text{SF}_\text{max}$ is more than 0.002 (Fig.~\ref{fig:fig_13}), which is significantly higher than the minimum limit set in section~3.
Firstly, this indicates the absence of the influence of small fluctuations in SF on the result, which is naturally expected with a small measurement error. Secondly, there are high confidence levels for the found SF maxima: all of them have $p=1$ except for one, for which $p=0.94$.
Fig.~\ref{fig:fig_13} (left panel) shows that the points obtained from the analysis of aperture photometry data and SAP fluxes fall into the general trend and, in almost half of the cases, approximately correspond to each other, while the points related to PDCSAP fluxes significantly dislodged.
According to aperture photometry, the values of $\tau_\text{v}$ range from $\approx 1/3$ to $<1$ of the duration of the corresponding interval (Fig.~\ref{fig:fig_13}, middle panel). The dependence of the maximum amplitude on the characteristic time of variability in the corresponding interval is also preserved (Fig.~\ref{fig:fig_13}, right panel).
It is interesting to note that according to aperture photometry data, the dependencies $\text{SF}_\text{max}(\tau)$ and $\text{mag}_\text{var}(\tau)$, with corresponding Pearson correlation coefficients $r$=0.97 and 0.99, are stronger and traced over a wider range of $\tau$ than according to data from PDCSAP and SAP fluxes.

\begin{figure}
\centering
\includegraphics[width=0.7\columnwidth]{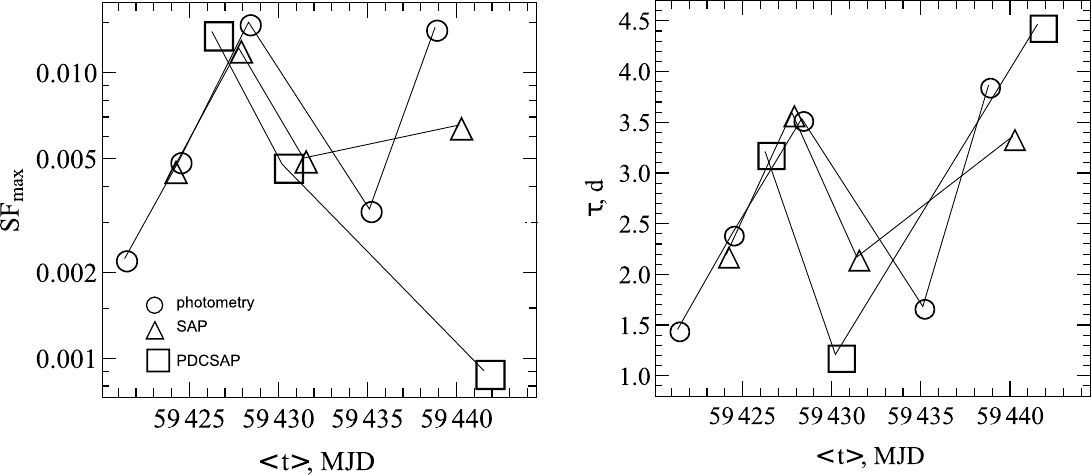}
\caption{Changing $\text{SF}_\text{max}$ and $\tau_\text{v}$ over time. PDCSAP and SAP data are provided for the corresponding light curves smoothed by a Gaussian with a core of 75 points.}
\label{fig:fig_14}
\end{figure}

Tracing the evolution of the average variability amplitude (Fig.~\ref{fig:fig_14}, left panel), it is interesting to note that the S5~1803+784  brightness weakening recorded in the ZTF data is reflected only in the decrease in $\text{SF}_\text{max}$ calculated by 600 s resolution data of PDCSAP fluxes.
There is no dependence of the characteristic time of variability on brightness (Fig.~\ref{fig:fig_14}, right panel), but it is noteworthy that $\tau_\text{v}$, obtained by different data types, agrees well.
Comparing the change in $\tau_\text{v}$ (Fig.~\ref{fig:fig_14}) and $\alpha$ (Fig.~\ref{fig:fig_3}) over time, in the range MJD~59427$-$59437, there is anticorrelation for SAP and aperture photometry data. However, a longer series of observations is needed for the final statement.

\section{Discussion}
\label{sec:Discuss}

The blazar brightness variability is often suggested to be described by a superposition of short-term fluctuations on long-term trends. For example, for blazar S5~0716+714, various mechanisms of short-term variability are assumed to operate on different time scales \cite{Rait21S5}.
The background for this was the different behavior of the color index with variability on different time scales. On the other hand, Gorbachev et al. \cite{Gorbachev22} note that in this case fine-tuning of the parameters and, possibly, the requirement of some synchronization of these mechanisms is required to ensure commensurate changes in the brightness of the object.
Additionally, Gorbachev et al. \cite{Gorbachev22} shown that the variability of S5~0716+714 at different time scales is explained within the framework of a single process --- in the presence of a curved radiation spectrum, the continuous appearance and evolution of sub-components (parts of the emitting region) with a different Doppler factor.
Since this can happen in most blazar jets, it is natural to expect indications of this variability mechanism for S5~1803+784. The sub-components can be of different volumes and Doppler factors, so it is not should expect a direct dependence of the characteristic time of variability on brightness.
Confirmation of the assumption is the change in the characteristic times of variability over time. Variability on long-time scales forms large-volume sub-components, whose abrupt change in the Doppler factor provides an increase/decrease in brightness on time scales of tens of days, while the evolution of smaller sub-components is responsible for a shorter-term variability.
Therefore, studies of short-term variability can shed light on the structure of the jet flow. This information is significant in terms of the acceleration of electrons and, possibly, protons at the boundary of layers at different speeds, which leads to an increase in radiation produced by inverse Compton scattering \cite{Ghisellini05} and the neutrinos formation \cite{Tavecchio14}.

With a power-law radiation spectrum, the brightness variability formed by a change in the Doppler factor is achromatic, while a change in the physical conditions in the emitting region leads to a change in $\alpha$.
On the other hand, data from long-term radio interferometric observations show that jets of active nuclei are curved and often have non-radially moving components \cite{Lister13, Lister21}. This fact indicates that there must be a variability in the jets caused by the $\delta$ change.
As a result of the analysis of 20 years of B-, V-, R-, and I-photometric observations of the blazar S5~0716+714, Gorbachev et al. \cite{Gorbachev22} concluded that the long-term variability of radiation originated from a change in $\delta$ with a non-power law spectrum arising from the action of synchrotron self-absorption.
The observations of blazar S5~1803+784 considered in this article cover an interval of $\sim$27 days. During this time, the blazar brightness gradually decreases by $\approx$0.5 magnitude, according to both ZTF data and our photometry of summarized TESS cuts.
During this, the spectral index undergoes variations with an amplitude of about 0.12, which does not depend on the object's brightness. Assuming that different physical parameters correspond to different spectral indices, it is difficult to naturally explain the smooth trend of decreasing the blazar brightness.

On the other hand, both the presence of sub-components in the jet flow and the curved radiation spectrum provide a simple explanation for the observed variability characteristics.
Deviations of the synchrotron radiation spectrum from the power law can be caused either by a break in the electron energy spectrum due to radiation losses or by the action of synchrotron self-absorption. In both cases, if $\delta$ is low, we observe a high-energy section of the electron radiation spectrum.
The greater the $\delta$, the lower the photon frequencies in the reference frame of a relativistically moving plasma correspond to the observed optical range, so the observed radiation spectrum becomes flatter.  Hence, if the long-term trend is caused by a decrease in the Doppler factor of the main part of the emitting region, then sub-components of different volumes and with different $\delta$ produce observed short-term brightness fluctuations, comparable with the object's total brightness, and characterized by a different $\alpha$.

Within this assumption of the variability formation, it follows that $\alpha$ depends only on $\delta$: a flatter radiation spectrum corresponds to a high $\delta$ of sub-components.
Additionally, the greater the $\delta$, the greater the average variability amplitude. Therefore, if the sub-components were of the same volume, a flatter spectrum and a shorter characteristic time with a larger variability amplitude should be expected.
However, from the comparison Fig.~\ref{fig:fig_3} and \ref{fig:fig_14} there is no strong correlation between $\text{SF}_\text{max}$ and $\tau_\text{v}$ from $\alpha$, therefore, the sub-components have different volumes.
The existing dependence of $\text{SF}_\text{max}$ on $\tau_\text{v}$ (Fig.~\ref{fig:fig_12} and \ref{fig:fig_13} left panel) we interpreted by the fact that the sub-components with a higher Doppler factor have a larger volume.

\section{Conclusion}
\label{sec:Concl}

In this article, we have proposed a new method for studying the short-term brightness variability of blazars. This method is applicable for a long, almost continuous series with a constant sampling rate.
The TESS data fulfills these requirements. The duration of one TESS sector is about 27 days, and objects located near the ecliptic poles have several consecutive observation sectors.
This fact opens up great opportunities for studying the short-term blazar variability. We have shown that for such faint objects, which are most blazars, TESS data in the form of SAP and PDCSAP fluxes cannot be applied directly due to the signal-to-noise ratio being less than one.
On the one hand, it would be possible to smooth out the noisy light curve for further analysis. For an instrumental TESS flux of $100-$200 counts per second, the best option is to use Gaussian convolution with a core of 75 points for data obtained with a time resolution of 600 seconds. Data with a time resolution of 120 seconds is much more noisy and is not inapplicable for analysis.
In general, the window size for smoothing may differ due to the object's brightness variations and different magnitudes for different objects. We propose to define this size as the smallest one for which the ``plateau'' on a plot $\text{SF}_{\max} \left( \tau\right)$ is absent.

On the other hand, an increase in measurement accuracy can be achieved by summing consecutive cuts of full-frame images. 
The performed aperture photometry for S5 1803+784 using comparison stars perfectly corresponds with the ZTF data, while PDCSAP fluxes are not.
The results based on the comparative aperture photometry data have the highest significance level, and we recommend uniquely using photometry based on summing up several pieces of TESS cuts to analyze the variability of blazars.

We applied a new method to study the temporal variability characteristics of the blazar S5~1803+784. The median of the significance levels of the found characteristic times of variability according to aperture photometry data exceeds $3\sigma$, while for TESS data with a time resolution of 600 seconds, the median is slightly more than $2\sigma$.
Therefore, the aperture photometry data and, to some extent, consistent data based on the 600-second light curve of SAP fluxes, convolved with a Gaussian with a core of 75 points, actually reflect the characteristics of the object's variability and are not related to random processes, noise in the data or affected by the limitations of the data series.
Note that our proposed method of analyzing the light curve is universal and can be applied to another series of observational data, the duration of which is much longer than the expected characteristic time of variability. For example, the Fermi-LAT space observatory provides suitable data for about a thousand blazars.
Note, that the more gaps in the data and the stronger the variability, the greater the distortion of the results of the time analysis.
Therefore, in the case of a noticeable data omission, additional statistical studies of the significance of the obtained results are necessary.

The properties of the blazar S5~1803+784 variability are very variable on time scales of several days. It may be explained by the alternating determining contribution to the observed flux of different parts of the emitting region with slightly different parameters.
That is, the characteristics of the observed short-term variability are the result of fluctuations in parameters with one active variability mechanism. Otherwise, it would be difficult to coordinate two or more variability processes to ensure comparable amplitudes and characteristic times.
Thus, we believe that the S5~1803+784 brightness variability is explained by the constant occurrence and evolution in the emitting region of sub-components having different volumes and Doppler factors. Under this, the radiation spectrum becomes flatter with decreasing frequency.

This method of studying the evolution of the average amplitude and the characteristic time of variability will later be applied to a longer series provided by TESS for the S5~1803+784 blazar. Perhaps, as the data increases, various dependencies for $\text{SF}_\text{max}$ and $\tau_\text{v}$ will appear.

\textit{\textbf{Acknowledgments}}. This work was partly supported by the Russian Science Foundation, No. 24-22-00343.
This paper includes data collected by the TESS missions and obtained from the MAST data archive at the Space Telescope Science Institute (STScI). 
Funding for the TESS mission is provided by the NASA Explorer Program.
Based on observations obtained with the Samuel Oschin Telescope 48-inch and the 60-inch Telescope at the Palomar Observatory as part of the Zwicky Transient Facility project. 
\\ZTF is supported by the National Science Foundation under Grants Numbers AST-1440341 and AST-2034437 and a collaboration, including current partners Caltech, IPAC, the Weizmann Institute of Science, the Oskar Klein Centre at Stockholm University, the University of Maryland, Deutsches Elektronen-Synchrotron and Humboldt University, the TANGO Consortium of Taiwan, the University of Wisconsin at Milwaukee, Trinity College Dublin, Lawrence Livermore National Laboratories, IN2P3, University of Warwick, Ruhr University Bochum, Northwestern University and former partnersthe University of Washington, Los Alamos National Laboratories, and Lawrence Berkeley National Laboratories. 
Operations are conducted by COO, IPAC, and UW. The ZTF forced-photometry service was funded under the Heising-Simons Foundation grant number 12540303 (PI: Graham).


\end{document}